\documentclass[letter,traditabstract]{aa}  

\usepackage{graphicx}
\usepackage[]{natbib}
\usepackage{url}
\usepackage{multirow}
\usepackage{txfonts}

\begin{document}

\newcommand{\uchile}{Departamento de Astronom\'ia, Universidad de Chile, 
Casilla 36-D, Santiago, Chile} 
\newcommand{\iap}{Universit\'e Paris 6, Institut d'Astrophysique de Paris, 
CNRS UMR 7095, 98bis bd Arago, 75014 Paris, France}
\newcommand{\iucaa}{Inter-University Centre for Astronomy and Astrophysics, 
Post Bag 4, Ganeshkhind, 411\,007 Pune, India}
\newcommand{\eso}{European Southern Observatory, Alonso de C\'ordova 3107, 
Vitacura, Casilla 19001, Santiago 19, Chile}

\newcommand{\avg}[1]{\left< #1 \right>} 

   \title{The evolution of the Cosmic Microwave Background Temperature\thanks{Based on observations carried out at the European Southern Observatory (ESO) using the Ultraviolet and Visual Echelle Spectrograph (UVES) at the 
Very Large Telescope (VLT, UT2-Kueyen) under Prgm.~IDs 278.A-5062(A), 081.A-03334(B), 082.A-0544(A), and 083.A-0454(A).}} 
  \subtitle{Measurements of $T_{\rm CMB}$ at high redshift from carbon monoxide excitation 
}

   \author{P. Noterdaeme\inst{1}
          \and
          P. Petitjean\inst{2}
          \and
          R. Srianand\inst{3}
          \and
          C. Ledoux\inst{4}
          \and
          S. L\'opez\inst{1}
          }

\institute{\uchile\\
\email{pasquier@das.uchile.cl, slopez@das.uchile.cl} 
\and \iap\\  
\email{petitjean@iap.fr} 
\and \iucaa\\ 
\email{anand@iucaa.ernet.in}
\and \eso\\  
\email{cledoux@eso.org}}


   \date{Received /Accepted}
 
   \abstract{
A milestone of modern cosmology was the prediction and
serendipitous discovery of the Cosmic Microwave Background (CMB), 
the radiation left over after decoupling from matter in the early 
evolutionary stages of the Universe.
A prediction of the standard hot Big-Bang model is the linear increase 
with redshift of the black-body temperature of the CMB ($T_{\rm CMB}$). 
This radiation excites the rotational levels of some interstellar molecules, 
including carbon monoxide (CO), which can serve as cosmic thermometers.
Using three new and two previously reported CO absorption-line systems 
detected in quasar spectra during a systematic survey carried out using 
VLT/UVES, we constrain the evolution of $T_{\rm CMB}$ to $z\sim3$. Combining 
our precise measurements with previous constraints, we obtain 
$T_{\rm CMB}(z)=(2.725\pm0.002)\times (1+z)^{1-\beta}$~K with 
$\beta=-0.007\pm0.027$, a more than two-fold improvement in precision. 
The measurements are consistent with the standard (i.e. adiabatic, $\beta=0$) 
Big-Bang model and  provide a strong constraint on the effective equation of 
state of decaying dark energy (i.e. $w_{eff}=-0.996\pm0.025$).}


   \maketitle


\section{Introduction}

The existence of the Cosmic Microwave Background (CMB) radiation is a 
fundamental prediction of the hot Big-Bang theory. If gravitation is 
described by general relativity and electromagnetism by Maxwell theory 
then photons propagate along null geodesics and the CMB black-body 
temperature must follow the relation 
$T_{\rm CMB}(z)=T_{\rm CMB}^0\times(1+z)^{1-\beta}$, 
with $\beta=0$ and where $T_{\rm CMB}^0$~=~2.725$\pm$0.002~K \citep{Mather99} 
is the temperature measured locally (at redshift $z=0$).
This relation, which is a theoretical consequence of the adiabatic expansion of the Universe, needs to be verified by direct measurements. 
This has also deeper theoretical implications \citep[][]{Uzan04}. 
A non-zero $\beta$ would indicate either a violation of the
hypothesis of local position invariance (and thus of the equivalence principle) 
or that the number of photons is not conserved -- with the constraint that 
the energy injection does not induce spectral distortion of the CMB.
In the first case, this should be associated with a variation of the 
fundamental constants \citep[see e.g.][]{Murphy03,Srianand04}.

There are currently two methods to measure $T_{\rm CMB}$ at redshifts $z>0$. 
The first one relies on the measurement of a small change in the spectral 
intensity of the CMB toward clusters of galaxies due to inverse Compton 
scattering of photons by the hot intra-cluster gas: the so-called 
Sunyaev-Zel'dovich (S-Z) effect \citep{Fabbri78,Rephaeli80}. Although this 
technique permits precise measurements \citep[$\Delta T \sim 0.3$\,K;][]{Battistelli02,Luzzi09}, 
the method is essentially limited to $z<0.6$ because of the scarcity of known clusters 
at higher redshifts. 
The other technique uses the excitation of interstellar atomic or molecular 
species that have transition energies in the sub-millimetre range and can be 
excited by CMB photons. 
When the relative population of the different energy levels are in radiative 
equilibrium with the CMB radiation, the excitation temperature of the species 
equals that of the black-body radiation at that redshift. 
Therefore, the detection of these species in diffuse gas, where collisional 
excitation is negligible, provides one of the best thermometers for 
determining the black-body temperature of the CMB in the distant Universe 
\citep[][]{Bahcall68}.  

\section{Constraining $T_{\rm CMB}(z)$ using QSO absorption systems}

The observation of diffuse gas at high redshift is routinely achieved through the 
absorption lines that the interstellar medium of high-$z$ galaxies produces in the 
spectra of bright background sources such as quasars. However, detecting suitable 
species for measuring $T_{\rm CMB}$ turns out to be difficult. 

\subsection{Atomic species}
Several attempts to measure $T_{\rm CMB}$ at high redshift have been made using 
the excitation of fine-structure levels of atomic species like carbon. 
These measurements generally led to upper-limits on $T_{\rm CMB}$ because collisional 
excitation dominates \citep[e.g.][]{Meyer86, Lu96, Roth99, Ge01}.
Assuming the CMB is the only source of excitation, \citet{Songaila94} measured 
$T_{\rm CMB}=7.4\pm0.8$~K from C$^0$ at $z=1.776$ toward the quasar Q\,1331$+$170. 
\citet{Cui05} later derived 
the physical conditions in the same system through the observation of molecular hydrogen 
\citep[see also][]{Carswell10}. 
Applying the inferred physical conditions to the excitation of C$^0$, they got 
$T_{\rm CMB}=7.2\pm0.8$~K. 
Similarly, \citet{Ge97} estimated $T_{\rm CMB}=7.9\pm1.0$~K at $z=1.9731$ 
from the excitation of C$^0$ toward Q\,0013$-$004. \citet{Srianand00} performed a detailed study 
of the physical conditions in the H$_2$-bearing system toward PKS\,1232$+$0815, which allowed for 
both stringent lower ($>6$\,K) and upper ($<14$\,K) limits on $T_{\rm CMB}$ at $z=2.3377$. 
Finally, \citet{Molaro02} used C$^+$ to constrain $T_{\rm CMB}=12.1^{+1.7}_{-3.2}$~K at $z=3.025$ 
toward Q\,0347$-$3819. 
In all these cases, collisional excitation was not negligible and had to be corrected for, 
introducing uncertainties in the measurement. 
Taking a conservative approach, \citet{Srianand08} assumed the CMB to be the only source of 
excitation to provide stringent upper-limits on $T_{\rm CMB}$ for a large sample of C\,{\sc i} 
absorption lines detected in high signal-to-noise, high-resolution spectra.
Other systematic effects might also be hard to control. For example, 
the accuracy of the oscillator strengths of C$^0$ fine-structure levels 
\citep[e.g.][]{Jenkins01,Morton03} 
is subject to discussions. 
It is therefore important to find a direct and independent way of measuring $T_{\rm CMB}$ 
at similar redshifts.

\subsection{Molecular species}
The rotational excitation of molecules with permanent dipolar momentum can provide 
a more direct and precise measurement. For instance, the CN molecule has proved to be a 
remarkable thermometer of the CMB in the Galaxy and has been used for precise measurement 
of $T_{\rm CMB}$ in different directions \citep[e.g.][]{Meyer85,Kaiser90,Roth93,Ritchey10}. However, up to now, CN 
has not been detected in diffuse gas at high redshift.
On the other hand, an exciting possibility is offered by carbon monoxide (CO) whose energy 
differences between rotational levels are close to $kT_{\rm CMB}$ at high redshift. The use 
of CO in this kind of experiments was prevented by the fact that CO has been detected 
in absorption only very recently at high-$z$ \citep{Srianand08, Noterdaeme09co, Noterdaeme10co}. 
Indeed, the detection of CO in diffuse gas at high redshift is challenging because of the low dust 
opacity required to observe the background source.
Because shielding from the surrounding UV radiation field becomes more efficient 
deeper inside an interstellar cloud, carbon is found predominantly in its ionised (C$^+$), atomic 
(C$^0$) and molecular forms (CO) in the external, intermediate, and deepest layers of the cloud, 
respectively. 
Since the lines of sight passing through the central and fully molecular part of the cloud 
are affected by strong extinction, we targeted the C$^0$ phase 
so that the light from the corresponding background quasar is only moderately obscured by the 
diffuse molecular gas while some CO is present in the gas. 

\section{Candidate search, follow-up, and analysis}
We have searched over 40\,000 quasar spectra from the Sloan Digital Sky Survey data base \citep[SDSS-DR7,][]{SDSSDR7}
for strong intervening C\,{\sc i} absorption lines that can be detected even in low resolution spectra. 
This led to the detection of 67 C$^0$-bearing absorption systems at redshifts between 1.5 and 3.
We note that no pre-selection was made upon the presence of neutral atomic hydrogen, 
i.e., the systems needed not be damped Lyman-$\alpha$ systems (DLAs, see \citealt{wolfe05})
\footnote{Indeed, several systems turn out to be sub-DLAs (i.e. with $N($H$^0)<2\times10^{20}$~cm$^{-2}$).}.
Indeed, the Lyman-$\alpha$ line is covered by the SDSS spectra only for $z_{\rm abs}\ge2.2$, 
while C\,{\sc i} can be detected down to $z_{\rm abs}\sim1.5$.

Because CO absorption lines are expected to be weak and not detectable at the SDSS 
spectral resolution ($\lambda/\Delta \lambda \simeq 2000$), the most promising candidates 
were subsequently 
re-observed at a resolving power of $\lambda/\Delta \lambda \simeq 50\,000$ 
with the Ultraviolet and Visual Echelle Spectrograph (UVES), 
mounted on the ESO Very Large Telescope. 
We present here three new detections of CO for which we derive the rotational 
excitation temperatures. Adding two previous measurements using CO 
by our group \citep{Srianand08, Noterdaeme10co}\footnote{Note that we 
have detected a sixth CO-bearing system \citep[at $z=1.64$ toward 
SDSS\,J160457.50+220300.5,][]{Noterdaeme09co}. However, this system is not 
suitable for constraining $T_{\rm CMB}$ because of the low S/N ratio 
achieved and the complexity of the absorption profile.}, we constrain the 
CMB black-body temperature at redshifts $1.7<z<2.7$.
Each quasar line of sight was observed for 
a total of three to nine hours with UVES. The data were reduced using the UVES 
pipeline, and individual spectra were shifted to the heliocentric-vacuum frame 
and combined together by weighting each pixel by the inverse of the associated variance.

The CO molecule possesses rotational, vibrational and electronic transitions 
which correspond to photons with, respectively, radio, infrared and far-ultraviolet wavelengths. 
We detect several absorption bands from the electronic transitions of CO in 
the vibrational ground-state, redshifted into the optical wavelength range. 
Each electronic band is split into three branches composed of several absorption lines 
arising from different rotational levels.
These absorption lines are resolved in the UVES data and the 
excitation temperatures of CO can be measured precisely thanks to 
the large number of lines (for each rotational level and 
spanning a range of oscillator strengths) that are fitted simultaneously. 

\begin{figure}[]
\renewcommand{\tabcolsep}{1.5pt}
\begin{tabular}{cc}
\includegraphics[height=0.58\hsize,bb= 42 325 503 758]{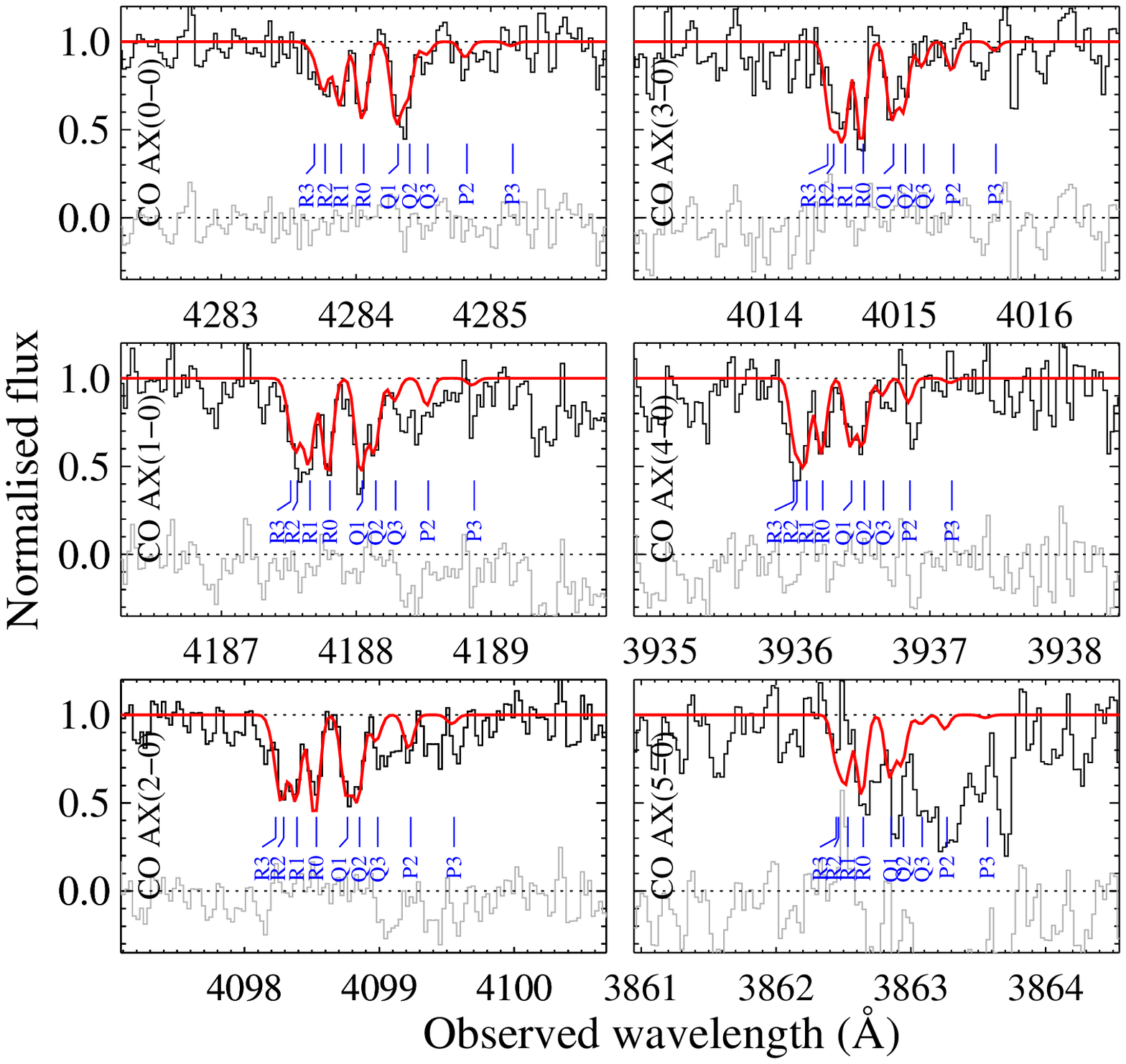}&
\includegraphics[height=0.58\hsize,bb= 37 340 308 758]{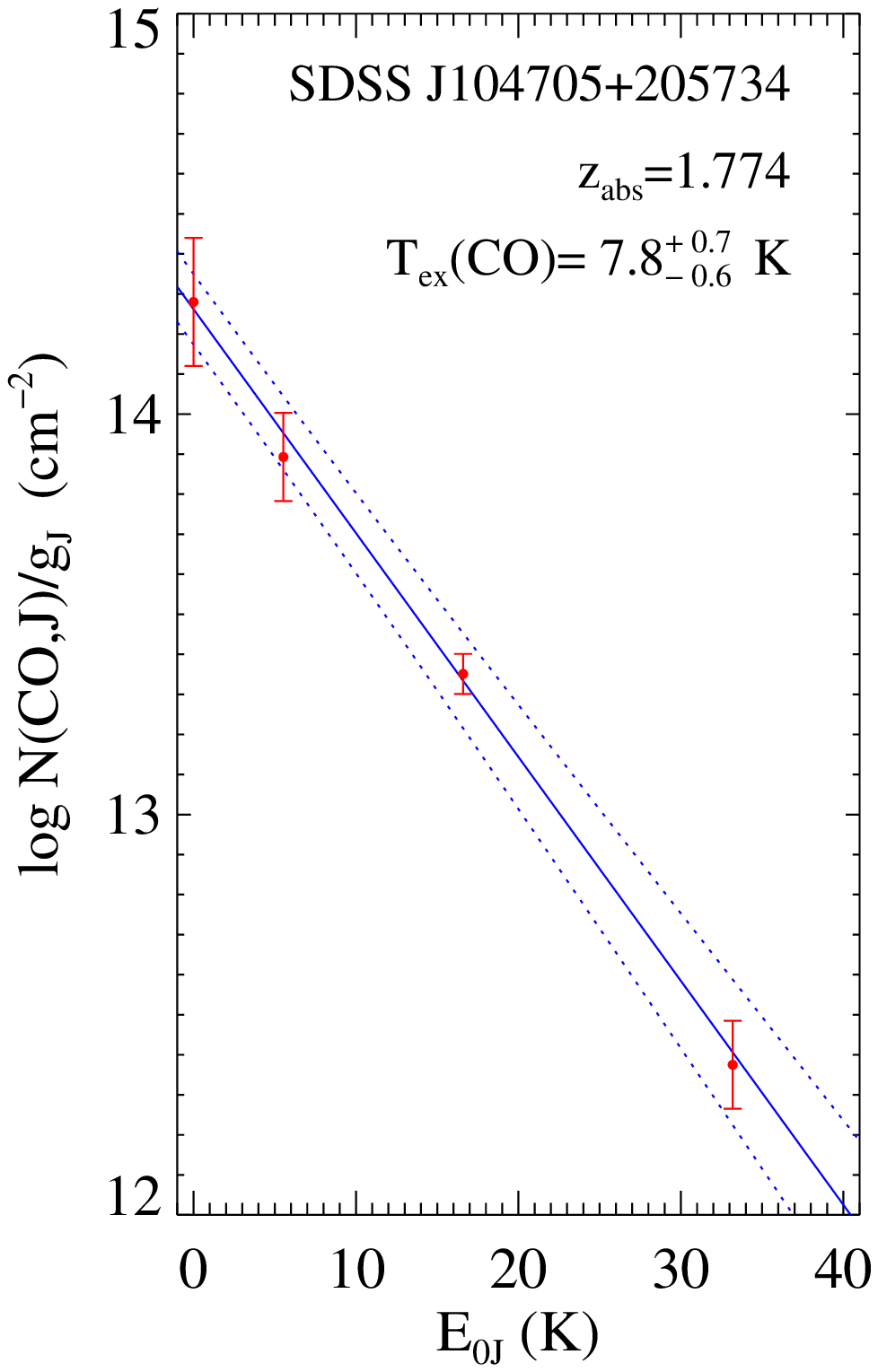}\\
\end{tabular}
\caption{Left: $A^1\Pi(\nu^\prime)\,\leftarrow\,X^1\Sigma^+(\nu$=$0)$ electronic bands of CO detected 
at $z_{\rm abs}=1.774$ toward the quasar SDSS\,J104705+205734. The modelled profile is over-plotted in red with the 
fitting residuals shown in grey.
Short vertical lines indicate the branch ('R', 'P' or 'Q') and the rotational level 
of the transition (J=0-3). Right: The corresponding excitation diagram of CO gives the column 
density of the rotational levels J weighted by their quantum degeneracy ($g_{\rm J}$) versus the energy of these 
levels relative to the ground state. 
The excitation temperature is directly given by $-1/(a \ln 10)$ where $a$ is the slope of the linear fit 
(solid blue line, with 1\,$\sigma$ errors represented by the dotted curves). 
\label{1047}}
\end{figure}

\begin{figure}[]
\renewcommand{\tabcolsep}{1.5pt}
\begin{tabular}{cc}
\includegraphics[bb=42 325 503 758,height=0.58\hsize]{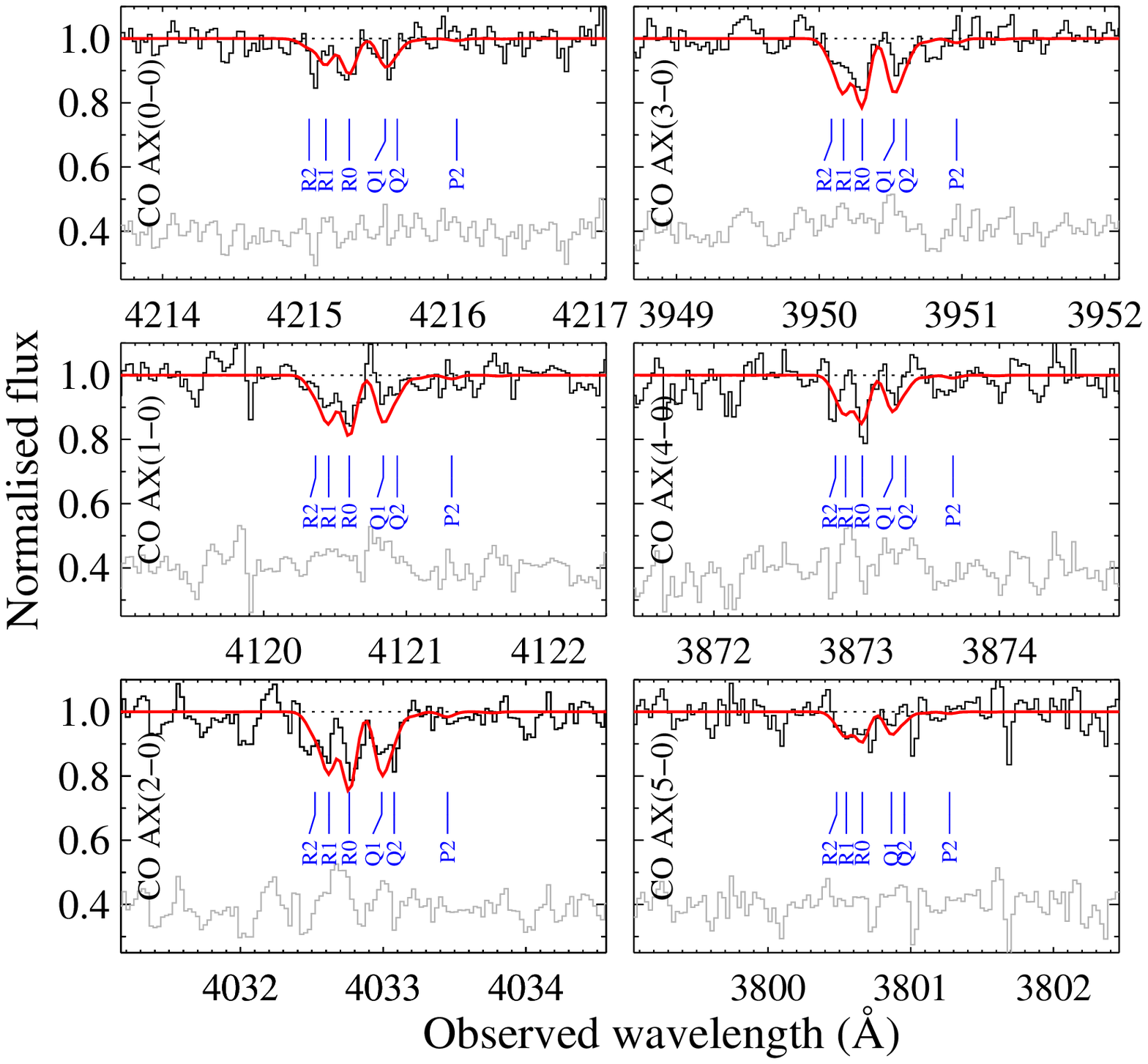}&
\includegraphics[height=0.58\hsize]{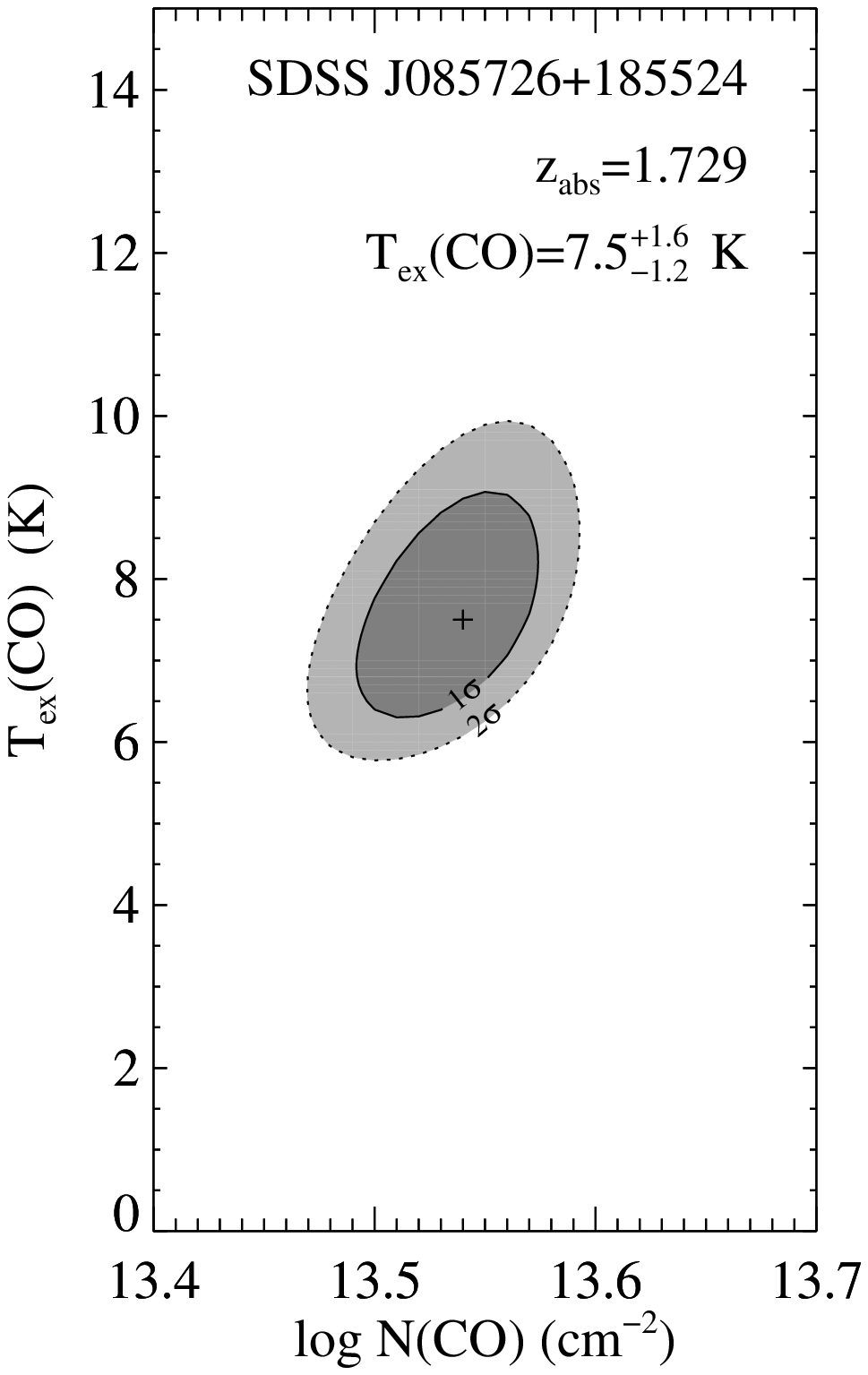}\\
\end{tabular}
\caption{Left: The CO electronic absorptions from the system at $z_{\rm abs}=1.729$ toward SDSS\,J085726+185524 are plotted together with the best-fit model 
($\log N$=13.54, $T_{\rm ex}$=7.5\,K, red profile).
Right: Confidence contours (solid:1\,$\sigma$, dotted:2\,$\sigma$) for the excitation temperature vs the total CO column density. \label{0857}}
\end{figure}

\begin{figure}[]
\renewcommand{\tabcolsep}{1.5pt}
\begin{tabular}{cc}
\includegraphics[bb=42 325 503 758,height=0.58\hsize]{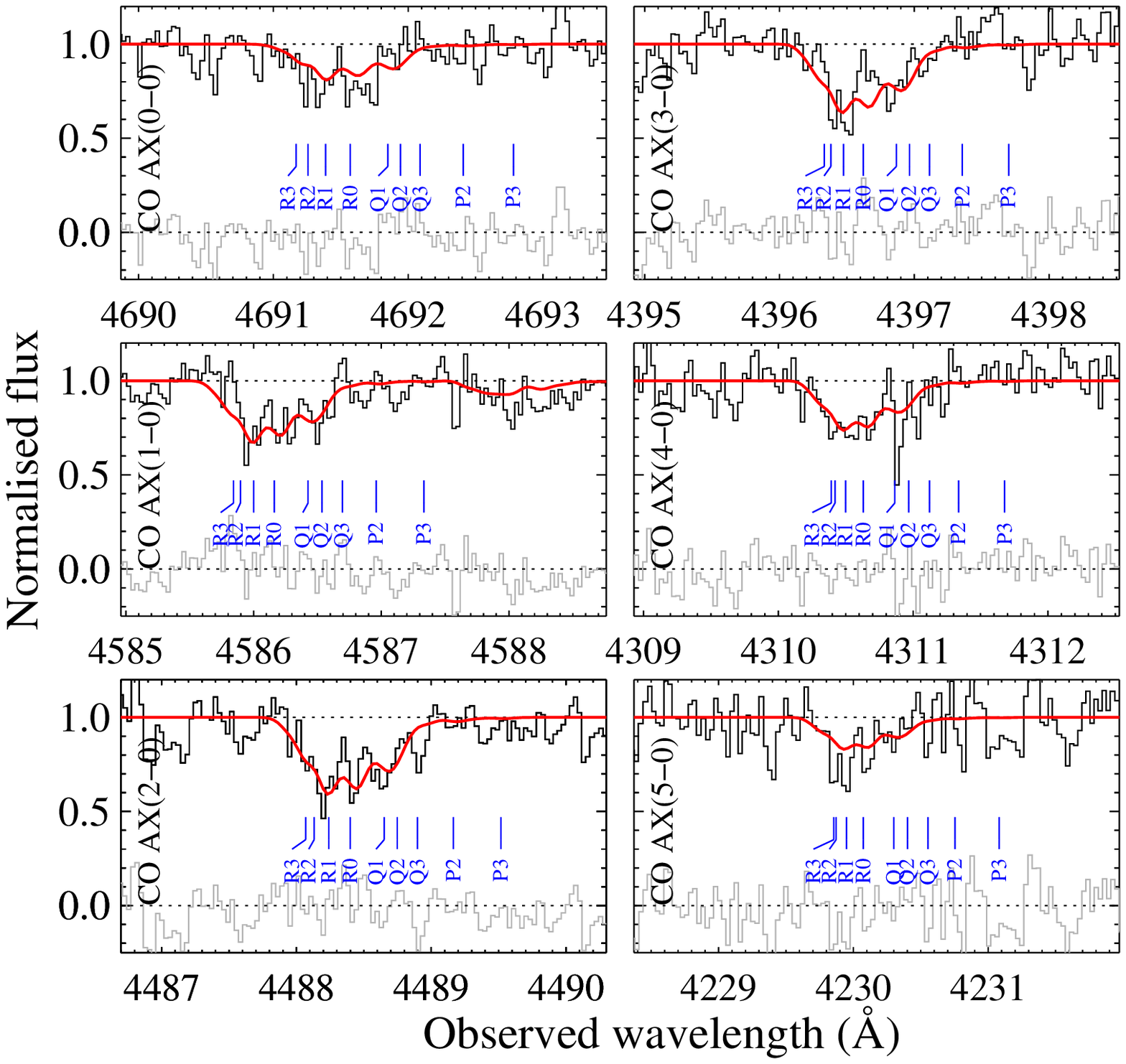}&
\includegraphics[height=0.58\hsize]{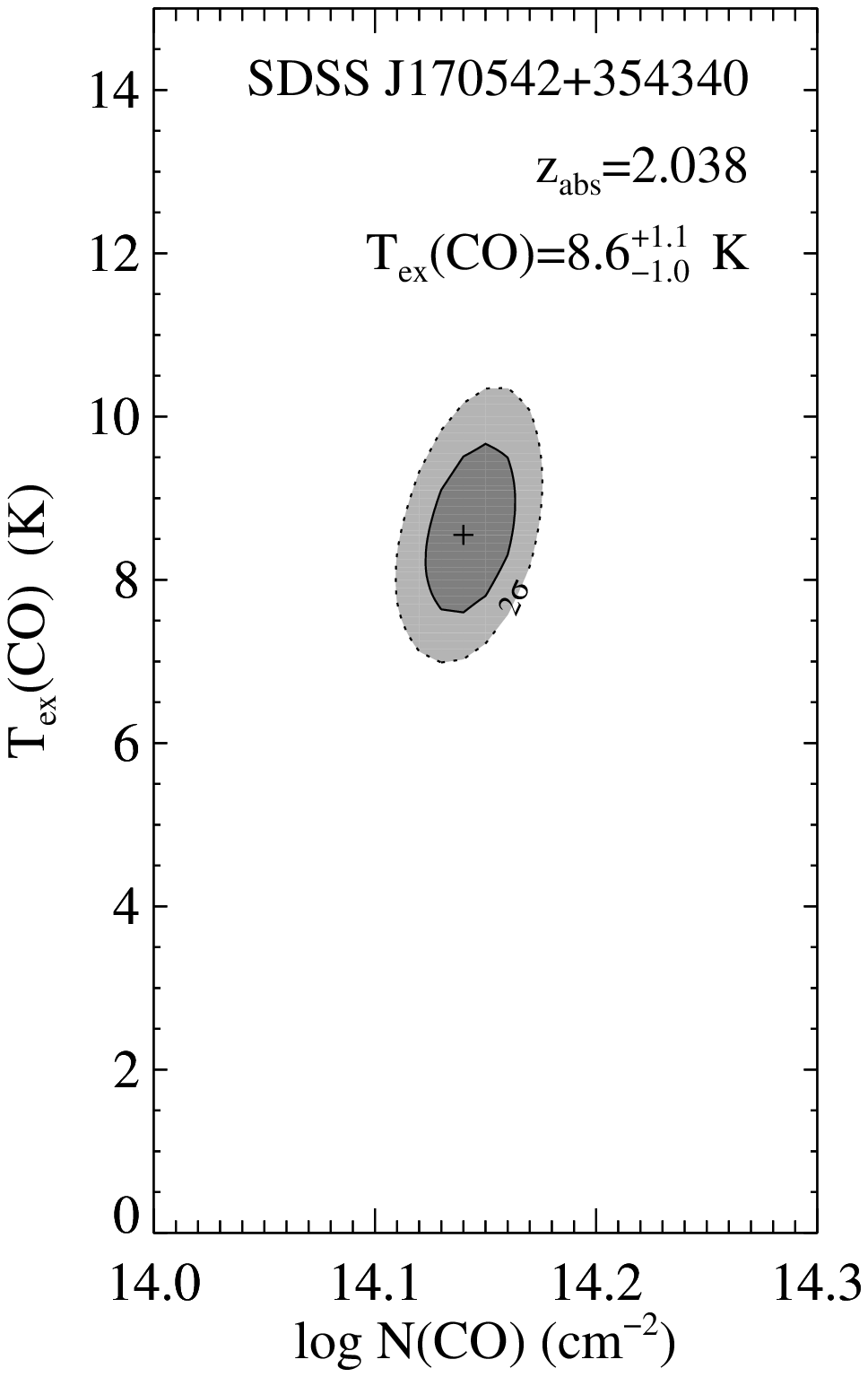}\\
\end{tabular}
\caption{Same as Fig.~\ref{0857} for the system at $z_{\rm abs}=2.038$ toward SDSS\,J170542+354340 
(best-fit model: $\log N$=14.14, $T_{\rm ex}$=8.6\,K).\label{1705}}
\end{figure}

The CO profiles were modelled with multiple Voigt profiles using a $\chi^2$-minimisation 
technique. The wavelengths and oscillator strengths 
were taken from \citet{Morton94}. 
For three of the five CO-bearing systems 
presented here, the profiles are well resolved and the signal-to-noise ratio 
obtained is sufficiently high to measure the column densities in rotational levels up to 
J~=~3 independently. 
The excitation temperature is then measured by modelling the rotational level populations 
with a single Boltzmann distribution as shown in the right-hand side panel of Fig.~\ref{1047}.
The detailed analysis of the systems toward SDSS\,J143912+111740 and SDSS\,J123714+064759 were 
presented in \citet{Srianand08} and \citet{Noterdaeme10co}, respectively. 
We used the same Voigt-profile fitting techniques to analyse the system toward 
SDSS\,J104705+205734 (see Fig.~\ref{1047}). A single velocity component was used to model the CO profile, 
leaving the column density in each rotational level, the Doppler parameter 
$b$, and the redshift of the CO-bearing cloud free. All absorption lines from the AX\,(0-0) 
to the AX\,(4-0) band were fitted simultaneously, while the AX\,(5-0) band was ignored 
during the fitting process due to the presence of a clear blending. We got 
$\log N($CO$)=14.74\pm0.07$, $b=1.1\pm0.1$~km\,s$^{-1}$ and $T_{\rm ex}($CO$)=7.8^{+0.7}_{-0.6}$~K. 
For the remaining two systems (toward SDSS J085726+185524 and SDSS\,J170542+354340), the 
rotational levels lines are hardly resolved\footnote{This is mainly due to larger Doppler 
parameters in these systems ($b\sim5$~km\,s$^{-1}$) similar to those of the S\,{\sc i}
 lines associated to the CO components.} and the signal-to-noise ratio is not high enough 
to measure the column densities 
in each rotational level independently. 
Instead, we used a technique similar to that presented in \citet{Burgh07} and \citet{Prochaska09}: a single 
excitation temperature was used as an external parameter to model the CO profile directly, 
again, using a Boltzmann distribution of the rotational level populations. The best fit-model 
was chosen from the minimum 
$\chi^2$ obtained for a grid of total column densities and excitation temperatures (see 
right-hand side panels of Figs.~\ref{0857} and \ref{1705}). 

The excitation temperatures for the five CO-bearing systems 
are given in Table~\ref{tco}.
At high redshift, the excitation of CO is dominated by radiative excitation as predicted 
in diffuse interstellar clouds \citep{Warin96,Burgh07}. %
Indeed, the excitation temperatures we measured are well above the mean 
temperature measured in the Galaxy for similar CO column densities \citep[$\avg{T_{\rm ex}}=3.6$~K, see][]{Burgh07}. 
This is further supported by the low volume density of the 
gas derived from the analysis of H$_2$ and C$^{\rm 0}$ lines in two of these high-$z$ 
absorption systems \citep{Srianand08, Noterdaeme10co}. 
Therefore, $T_{\rm ex}$(CO) must be a good proxy for $T_{\rm CMB}$ at high redshift.

\begin{table}
\caption{CO excitation temperatures. \label{tco}}
\centering
\begin{tabular}{cccc}
\hline
\hline
Quasar & $z_{\rm abs}$ & $T_{\rm ex}$(CO)~(K) & Ref. \\
\hline
{\large \strut} SDSS\,J085726+185524 & 1.7293 & 7.5$^{+1.6}_{-1.2}$  & 1 \\
{\large \strut} SDSS\,J104705+205734 & 1.7738 & 7.8$^{+0.7}_{-0.6}$  & 1 \\
{\large \strut} SDSS\,J123714+064759 & 2.6896 & 10.5$^{+0.8}_{-0.6}$ & 2 \\
{\large \strut} SDSS\,J143912+111740 & 2.4184 & 9.15$^{+0.7}_{-0.7}$ & 3 \\
{\large \strut} SDSS\,J170542+354340 & 2.0377 & 8.6$^{+1.1}_{-1.0}$  & 1 \\
\hline
\end{tabular}
\tablebib{(1) this work; (2) \citet{Noterdaeme10co}; (3) \citet{Srianand08}}
\end{table}

\section{Conclusion}

\begin{figure*}
\centering
\includegraphics[bb=24 124 566 396,clip=, width=0.9\hsize]{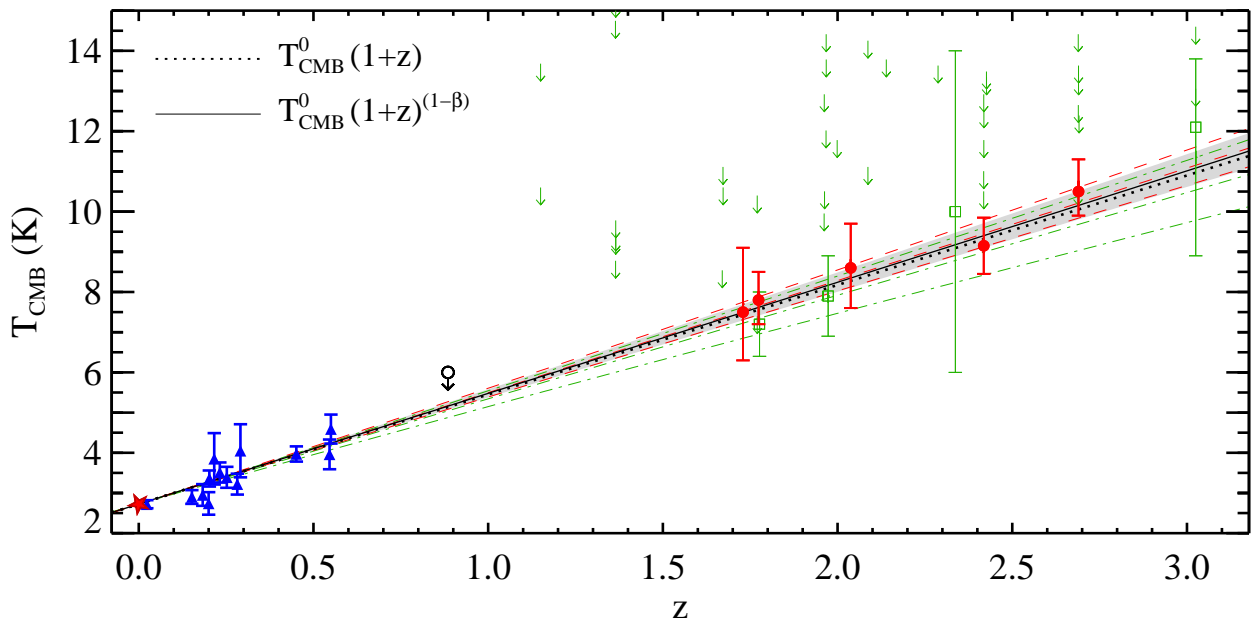}
\caption{The black-body temperature of the Cosmic Microwave Background radiation as a 
function of redshift. The star represents the measurement at $z=0$ \citep{Mather99}. 
Our measurements based on the rotational excitation of CO molecules are represented 
by red filled circles at $1.7<z<2.7$. Other measurements at $z>0$ are based 
(i) on the S-Z effect (blue triangles at $z<0.6$, \citealt{Luzzi09}) 
and (ii) on the analysis of the fine structure of atomic carbon (green open squares: 
$z=1.8$, \citealt{Cui05}; $z=2.0$, \citealt{Ge97}; $z=2.3$, \citealt{Srianand00}; 
$z=3.0$, \citealt{Molaro02}). 
Upper-limits come from the analysis of atomic carbon (from the literature and our UVES sample, see 
\citealt{Srianand08}) and from the analysis of molecular absorption lines in the lensing galaxy 
of PKS\,1830-211 \citep[open circle at $z=0.9$,][]{Wiklind96}.
The dotted line represents the adiabatic evolution of $T_{\rm CMB}$ as expected in standard 
hot Big-Bang models. The solid line with shadowed errors is the fit using all the data 
and the alternative scaling of $T_{\rm CMB}(z)$ \citep{Lima00} yielding $\beta=-0.007\pm0.027$. 
The red dashed curve (resp. green dashed-dotted) represents the fit and errors using S-Z + CO 
measurements (resp. S-Z + atomic carbon). \label{TCMBz}}
\end{figure*}

The CMB temperature derived from the rotational excitation of CO in five absorption systems 
(three studied here plus two previously published, see Table~\ref{tco}) 
are presented in Fig.~\ref{TCMBz} together with measurements and upper-limits 
obtained from the analysis of the populations of the fine-structure energy levels of atomic carbon 
or from the S-Z effect in galaxy clusters. 
An upper-limit has also been obtained from the analysis of millimetre absorption 
lines from different molecules in the gravitational lens of PKS\,1830-211 \citep{Wiklind96}. 

The technique presented here allows us to probe the temperature of the CMB at high redshift, 
providing constraints that are independent from and stronger than 
those arising from the analysis of C$^0$ and C$^+$.
This demonstrates that the rotational excitation of interstellar CO molecules can provide a 
direct and precise measurement of $T_{\rm CMB}$ in the early Universe. 

Fitting the measurements from different techniques with the expression 
$T_{\rm CMB}(z)=T_{\rm CMB}^0\times(1+z)^{1-\beta}$ \citep{Lima00}, we get 
the constraints on $\beta$ summarised in Table~\ref{beta}. 
We note that combining CO measurements of $T_{\rm CMB}(z)$ with those 
obtained from other techniques, improves the precision of the $\beta$-measurement 
by more than a factor of two.
The measurement presented here, $\beta=-0.007\pm0.027$, directly supports 
the adiabatic evolution of the CMB radiation temperature ($\beta=0$), expected from 
the standard hot Big-Bang model. 
Considering alternative $\Lambda$ cosmological models, \citet{Jetzer10} 
demonstrated that measuring $T_{\rm CMB}$ at 
different redshifts allows one to constrain the effective equation of state of 
decaying dark energy ($p=w_{eff} \rho$). 
Fitting the measurements of $T_{\rm CMB}$ with their temperature-redshift relation 
(Eq. 22 in \citealt{Jetzer10}), taking $\Omega_m=0.275\pm0.015$ \citep{Komatsu10} and 
fixing $\gamma$ to the canonical value (4/3), we get $w_{eff}=-0.996\pm0.025$ which is a 
tighter constraint compared to those previously derived from other methods 
\citep[e.g.][]{Kowalski08,Riess09,Kessler09,Jullo10}.

Finally, large and deep QSO surveys 
such as SDSS~III should provide more lines of sight along which CO can 
be detected while high-resolution spectrographs on future extremely large telescopes 
will allow for full de-blending of the CO lines in different rotational levels, 
yielding more accurate measurements.

\begin{table}
\caption{Constraints on the evolution of $T_{\rm CMB}$ with redshift \label{beta}}
\centering
\begin{tabular}{cc}
\hline \hline
Data set                 & $\beta$             \\
\hline
S-Z                      & $+0.040\pm0.079$    \\
S-Z + atom. carbon       & $+0.029\pm0.053$    \\
S-Z + CO                 & $-0.012\pm0.029$    \\
S-Z + atom. carbon + CO  & $-0.007\pm0.027$    \\
\hline
\end{tabular}
\end{table}

\acknowledgements{
We thank the referee, Paolo Molaro, for insightful comments that 
improved the paper. We are grateful to Jean-Philippe Uzan and Andrew 
Fox for their helpful comments on early versions of this paper.
We acknowledge the use of the Sloan Digital Sky Survey. 
PN is a CONICYT/CNRS fellow and acknowledges the ESO Office for Science 
for hospitality and support. PPJ and RS gratefully acknowledge support 
from the Indo-French Centre for the Promotion of Advanced Research under 
program N.4304-2. SL is supported by FONDECYT grant No. 1100214.}

\bibliographystyle{/scisoft/share/texmf/aa/aa-package/bibtex/aa}

\end{document}